\documentclass[12pt]{article}
\usepackage{amsmath}
\usepackage{amssymb}
\usepackage{amsthm}
\usepackage{amscd}
\usepackage{amsfonts}
\usepackage{graphicx}%
\usepackage{fancyhdr}
\usepackage{subfigure}
\usepackage[top=1in, bottom=1in, left=1.25in, right=1.25in]{geometry}
\newcommand{\mylistbegin}{
  \begin{list}{$\bullet$}
   {
     \setlength{\itemsep}{-2pt}
     \setlength{\leftmargin}{1em}
     \setlength{\labelwidth}{1em}
     \setlength{\labelsep}{0.5em} } }

\newcommand{\mylistend}{
   \end{list}  }
   \newcommand{\eg}{\textit{e.g.}}

\newcommand{\ie}{\textit{i.e.}}

\newcommand{\data}{\textbf{EgoNet-UIUC}}
\newcommand{\networkzip}{\textbf{EgoNetUIUC-LinkedinCrawl-Aug2013-Network.zip}}
\newcommand{\profilezip}{\textbf{EgoNetUIUC-LinkedinCrawl-Aug2013-Profiles.zip}}
\newcommand{\relationshiptxt}{\textit{relationship.txt}}
\newcommand{\labeltxt}{\textit{label.txt}}

\begin{document}
\vspace{-30mm}
\title{EgoNet-UIUC: A Dataset For Ego Network Research}
\date{}
\author{Rui Li, \space Kevin Chen-Chuan Chang
\\
\{ruili1,  kcchang\}@illinois.edu
\\
Computer Science Department,  \\
University at Illinois at Urbana-Champaign
}
\maketitle

\section{Overview}

In this report, we introduce the version one of \data, which is a dataset for ego-network research. In literature, an ego-network is defined as a network around an individual node, which contains relationships between the individual node and its neighbors and the relationships among the neighbors. Specifically, we collected the 230 ego networks from Linkedin during 2013. In total, there are 33K users (with their attributes) and 283K relationships (with their relationship types). We name the dataset as \data, which stands for  \textbf{Ego} \textbf{Net}work Dataset from \textbf{U}niversity of \textbf{I}llinois at \textbf{U}rbana-\textbf{C}hampaign.

In this report, we will explain how we collect \data\ in Section~\ref{sec-creation}, describe \data\ in details in Section~\ref{sec-description}, and provide instructions about how to obtain \data\ in Section~\ref{sec-usage}.

\section{Creation} \label{sec-creation}

\begin{figure}
  \center
  \includegraphics[height=40mm]{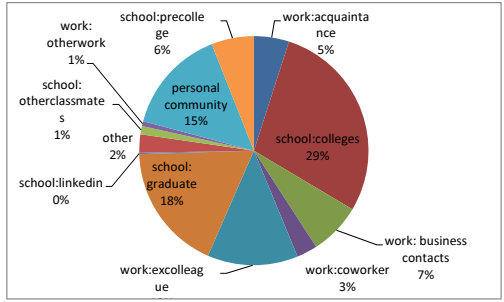}\\
  \vspace{-3mm}
  \caption{Percentages of Different Types of Relationships}\label{fig:example}
\end{figure}

\data\ (version 1) was originally collected from May to Aug 2013. We conducted a research study\footnote{http://forward.cs.illinois.edu/demos/linkedin/about.html} to collect the data. Initially, we invited a set of seed users to participate in our study online. Those users can further invite their friends to join. If a user participates our study, we collect his attributes (including \textit{Location}, \textit{Education}, and \textit{Work}) and his ego network (including the relationships from the ego user to his friends and  the relationships among his friends) based on Linkedin APIs, and we also ask the ego user to label his relationships with his friends from 11 categories in three domains (Personal Community, Work and School). Figure~\ref{fig:example} shows the different categories of relationship types and their percentages in the dataset.

By the end of the study, we collected about 230 users' ego networks. There are about 33K users with attributes (\ie, 230 users are linked to 33K friends in total), and 283K relationships, which include 1) 33K relationships from ego users to their friends and 2) 250K relationships among their friends. Among the 33K relationships between ego users and their friends, 15K relationships are labeled. To protect participants' privacy, we anonymized all users' attributes and their relationship labels by transferring them to ids.

\section{Description} \label{sec-description}
Now, we describe \data\ in details. Specifically, the dataset contains the following two zip files.
\begin{itemize}
  \item \networkzip\ contains 280K user relationships and 15K relationship labels,
  \item \profilezip\ contains different attributes of 30K users.
\end{itemize}

{\flushleft{\networkzip}} contains two txt files, \relationshiptxt\ and \labeltxt. \textit{relationship.txt} contains 280K relationships from 230 ego networks. Each relationship is a connection from a user to another user in an ego network. Figure~\ref{fig:relaitonship} illustrates its format with a concrete example. In the example, a record ``U0 U0 U1" means that, in U0's ego network, U0 connects to U1. \labeltxt\ contains labels for 15K relationships between ego users' and their friends. Each relationship is labeled by the ego user from the categories in Figure~\ref{fig:example}. The label (\eg, school:graduate) indicates the senmantic of the relationship (\eg, they are classmates in their grapduate school). We emphasize that the labeled relationships only contain the connections between the ego users and their friends.  Figure~\ref{fig:label} illustrates its format with a concreate example. Here, a record ``U10305	U10369	C2'' means that the type of relationship between U10305 and U10351 is C2.

\begin{figure}
\center
\begin{tabular}{|l|l|}
\hline
  Format & Example\\
  \hline
 EgoUserID1 tab FromUSERID  tab ToUserID & U0 U0	U1\\
 EgoUserID1 tab FromUSERID  tab ToUserID & U0	U27	U23\\
 .... & ... \\
 EgoUserID2 tab FromUSERID  tab ToUserID & U1046	U1046	U1172
\\
 EgoUserID2 tab FromUSERID  tab ToUserID & U1046	U1046	U1173\\
  \hline
\end{tabular}
 \vspace{-3mm}
  \caption{Format for \relationshiptxt}\label{fig:relaitonship}
\end{figure}

\begin{figure}
\center
\begin{tabular}{|l|l|}
\hline
  Format & Example\\
  \hline
 EgoUserID tab FriendID tab LabelID  & U3297	U3299	C4\\
 EgoUserID tab FriendID tab LabelID  & U10305	U10351	C2\\
  \hline
\end{tabular}
 \vspace{-3mm}
  \caption{Format for \labeltxt}\label{fig:label}
\end{figure}
\vspace{-1mm}
{\flushleft{\profilezip}} contains three txt files: \emph{location.txt}, \emph{education.txt}, and \emph{position.txt}. Each file records users' attribute values for a particular attribute (\eg, location). The attribute associated with a file is indicated by the file name (e.g., location).
A user may have multiple values for an attribute (\eg, a user may have multiple occupations). Figure~\ref{fig:education} illustrate the format of one of the files with examples. In the figure, U346 is UserID, 2 is the number of values associated with the attribute (\eg, education), and E0 is an attribute value.

\begin{figure}
\center
\begin{tabular}{|l|l|}
\hline
  Format & Example \\
  \hline
 USERID 1 & U346 \\
Number of Attribute Values & 2 \\
Attribute Value 1 & E0 \\
Attribute Value 2 & E1 \\
  ... & ... \\
  USERID 2 &  U23221 \\
 Number of Attribute Values  & 1\\
  Attribute Value 1  & E2 \\
  \hline
\end{tabular}
 \vspace{-3mm}
  \caption{Format for \emph{education.txt}}\label{fig:education}
\end{figure}

\section{Usage} \label{sec-usage}
\data\ is created for research purpose only. To obtain and use the dataset, you must agree with the following rules.
\begin{itemize}
 \item Use the data only for research.
 \item Not distribute the data to others.
 \item Participate in the research study\footnote{http://forward.cs.illinois.edu/demos/linkedin/about.html}.
 \item Cite this report appropriately in publications.
\end{itemize}

If you agree with the above rules, please send an email request to Rui Li (ruili1@illiois.edu). In the email, please clearly identify yourself (with your name and organization), and clearly state that you agree with the above rules. We need your name just for tracking the distribution of the dataset.

If you have any additional questions about the dataset, please email Professor Kevin Chang, or his student Rui Li. We also maintain an online description\footnote{https://wiki.engr.illinois.edu/display/forward/Dataset-CP-LinkedinCrawl-Aug2013} for this dataset.

\end{document}